\documentclass[aps,prl,superscriptaddress,twocolumn,10pt]{revtex4-1}
\usepackage{graphicx}
\usepackage{physics}
\usepackage{algpseudocode}
\usepackage{pythonhighlight}
\usepackage{amsmath,amssymb,amsfonts}
\usepackage{hyperref}
\usepackage{xcolor}
\usepackage{natbib}
\usepackage{changes}
\usepackage[english]{babel}
\usepackage{subcaption}
\usepackage{subfiles}
\usepackage{float}
\usepackage{siunitx}

\definecolor{Blue}{HTML}{1F77B4}
\definecolor{BlueLight}{HTML}{74B8FB}
\definecolor{BlueDark}{HTML}{04456E}
\definecolor{Red}{HTML}{D62728}
\definecolor{RedLight}{HTML}{FD9383}
\definecolor{RedDark}{HTML}{84030F}
\definecolor{Teal}{HTML}{037E8A}
\definecolor{Violet}{HTML}{7268AB}
\definecolor{NearBlack}{HTML}{141414}
\definecolor{Midgray}{HTML}{6B6B6B}
\definecolor{Lightgray}{HTML}{F2F2F2}

\newcommand{\ibm}{\texttt{ibm\_aachen}}
\newcommand{\ionq}{\texttt{Forte Enterprise}}

\AtBeginDocument{\RenewCommandCopy\qty\SI}

\begin{document}

\title{Large circuit execution for NMR spectroscopy simulation on NISQ quantum hardware}
\author{Artemiy Burov}
\affiliation{School of Life Sciences, University of Applied Sciences Northwestern Switzerland (FHNW), Hofackerstrasse 30, CH-4132 Muttenz, Switzerland}
\affiliation{\'{E}cole Polytechnique F\'{e}d\'{e}rale de Lausanne (EPFL), CH-1015 Lausanne, Switzerland}
\affiliation{MatterDecoder, CH-3008 Bern, Switzerland}
\author{Julien Baglio}
\email{julien.baglio@unibas.ch}
\affiliation{Center for Quantum Computing and Quantum Coherence (QC2), University of Basel, Klingelbergstrasse 82, CH-4056 Basel, Switzerland}
\affiliation{QuantumBasel, Schorenweg 44B, CH-4144 Arlesheim, Switzerland}
\author{Cl\'{e}ment Javerzac}
\email{clement.javerzac@fhnw.ch}
\affiliation{School of Life Sciences, University of Applied Sciences Northwestern Switzerland (FHNW), Hofackerstrasse 30, CH-4132 Muttenz, Switzerland}
\affiliation{\'{E}cole Polytechnique F\'{e}d\'{e}rale de Lausanne (EPFL), CH-1015 Lausanne, Switzerland}
\affiliation{MatterDecoder, CH-3008 Bern, Switzerland}

\begin{abstract}
With the latest advances in quantum computing technology, we are gradually moving from the noisy intermediate-scale quantum (NISQ) era characterized by hardware limited in the number of qubits and plagued with quantum noise, to the age of quantum utility where both the newest hardware and software methods allow for tackling problems which have been deemed difficult or intractable with conventional classical methods. One of these difficult problems is the simulation of one-dimensional (1D) nuclear magnetic resonance (NMR) spectra, a major tool to learn about the structure of molecules, helping the design of new materials or drugs. Using advanced error mitigation and error suppression techniques from Q-CTRL together with the latest commercially available superconducting-qubit quantum computer from IBM and trapped-ion quantum computer from IonQ, we present the quantum Hamiltonian simulation of liquid-state 1D NMR spectra in the high-field regime for spin systems up to 34 spins. Our pipeline has a major impact on the ability to execute deep quantum circuits with the reduction of quantum noise, improving mean square error by a factor of 22. It allows for the execution of deep quantum circuits and obtaining salient features of the 1D NMR spectra for both 16-spin and 22-spin systems, as well as a 34-spin system, which lies beyond the regime where unrestricted full Liouville‑space simulations are practical (32 spins, the Liouville limit). Our work is a step toward near-term quantum utility in NMR spectroscopy.
\end{abstract}

\maketitle

\section{Introduction\label{Introduction}}

The spin dynamics of \textit{nuclear magnetic resonance} (NMR) is a practical, real-world use case in which it is hoped to witness the first hints of quantum utility. NMR experiments are used, among many other things, in material sciences and pharmaceutical investigations to find new materials, catalysts~\cite{ZHENG20241} and drugs~\cite{LI2021116152} or for the study of biological compounds~\cite{AGGARWAL2022237}. An NMR spectrum is the result of a Fourier transform of some physical quantity associated with the molecule, for example, the total magnetization of the nuclear spins for a \textit{one-dimensional} (1D) spectrum. The quantum dynamics of a nuclear spin system is governed by a Heisenberg Hamiltonian for which the mathematical degrees of freedom are the same as the building blocks of a quantum circuit: Pauli operators. This explains why NMR spectroscopy is a promising use case to witness quantum utility, as there is a one-to-one correspondence between the quantum dynamics of the physical system and its simulation with a quantum computer. The numerical simulation of an NMR experiment is also known to be challenging on classical computers, in particular when the dynamics is hard because of strong spin-spin coupling interactions. There have been studies exploring quantum simulations for NMR spectroscopy in the low-field regime or solid-state NMR~\cite{demler1, demler2, google_obrian}, but very few for high-field liquid-state NMR~\cite{burov2024}. A review of the current state of the art can be found in~\cite{das2025}. The classical computing power required for NMR simulation grows exponentially with the size of the spin system and is limited to around 20 spins for a full classical simulation of the quantum dynamics~\cite{EDWARDS2014107}. This limit was also highlighted in high-field NMR simulations when comparing the scaling of classical computations with that of quantum simulations~\cite{burov2024}, and was also studied to identify which molecules are promising for quantum simulations~\cite{fratus2025}. This limit at around $N=20$ spins was raised to $N=32$ spins when taking into account advanced classical methods based on tensor networks, such as matrix product states~\cite{elenewski2025}. We refer to this limit as the \textit{Liouville limit} of NMR spectrum simulations. Our classical benchmark uses restricted Liouville space (coherence‑order cutoff and symmetry) as it is implemented in the SPINACH package~\cite{spinach} to make an $N=34$ restricted classical simulation tractable.

The most common technique to simulate NMR spectra using quantum computers relies on the quantum Hamiltonian evolution of the \textit{free induction decay} (FID) signal generated by the non-equilibrium nuclear spin magnetization. The computation can be performed either in the Hilbert space or in the Liouville space using density matrices. Trotterization of the time evolution~\cite{trotter1959product,childs2021} of the initial quantum state is performed for $K$ time steps, and for each of these steps, the expectation value of the magnetization operator is calculated. The FID signal is the collection of the $K$ expectation values, from which the Fourier transform is taken to obtain the 1D NMR spectrum. It is immediately clear that the quality of the quantum hardware on which this computation is performed will have a significant impact on the obtained spectrum, especially in the current \textit{noisy intermediate scale quantum} (NISQ) computing era. This impact has been mentioned and studied, e.g. in~\cite{burov2024, khedri2024} and was highlighted as one of the most important limitations of NISQ simulations. The impact of quantum noise grows with the depth of the quantum circuits used to perform the Hamiltonian evolution. The depth is, in turn, correlated with the complexity of the simulated molecule: with more spin systems and strong spin-spin couplings, using advanced error mitigation and error suppression techniques, together with the latest available quantum hardware with reduced two-qubit gate errors, becomes unavoidable.

We present in this work the quantum Hamiltonian simulation of three different molecules (nuclear spin systems) which have been identified as classically hard spin systems~\cite{demler2}\footnote{Also identified by Ilya Kuprov, lead developer of SPINACH~\cite{spinach}, the standard classical tool for NMR spectrum simulations. We also discussed this matter in personal communications with him as well as with the HQS Quantum Simulation company.}: the anti-3,4-difluoroheptane (coined DFH in the following) with 16 spins for which we present the simulation of the 1D $^{19}\text{F}$ NMR spectrum, a symmetric P-H molecule with two coupled tert-butyl groups with 22 spins for which we present both the simulations of the 1D $^1\text{H}$ and the 1D $^{31}\text{P}$ NMR spectra (coined symm\_H and symm\_P, respectively), and finally the B[ACR9]3 phosphorous system (also coined ``phosphorous cluster'') with 34 spins, beyond the Liouville limit, for which we present the 1D $^{31}\text{P}$ NMR spectrum. We employ Trotterization of the NMR Hamiltonian, with one Trotter step (one repetition). Each qubit corresponds to one nuclear spin from the molecule that we simulate.

Our computations have been performed primarily on the 156-qubit superconducting-qubit quantum computer \ibm\ provided by IBM. To investigate a different quantum hardware modality and in particular the effect of all-to-all connectivity on the results, we have also performed a computation of the symm\_P 1D $^{31}\text{P}$ NMR spectrum on the \ionq\ trapped-ion quantum computer from IonQ. Both types of computations have employed an advanced error suppression pipeline based on the Q-CTRL tool Fire Opal~\cite{fireopal,Mundada2023}, which has allowed for a substantial reduction of the actual depth of the quantum circuit on the quantum devices with low (nearly zero) overhead, together with between 12x and 22x improvement in accuracy of the simulation of the FID with respect to the raw output from the quantum devices. Note that other error suppression and mitigation pipelines also exist~\cite{lee2023,aharonov2025importance,aharonov2025reliable} and their exploration could also be interesting, such as QESEM from Qedma~\cite{aharonov2025importance,aharonov2025reliable}. Our simulations, when compared to restricted classical simulations from SPINACH~\cite{spinach}, reproduce the salient features of the 1D spectra. We will also present the limitation of our framework and propose a way to improve the results even further, paving the way for quantum utility in NMR spectroscopy simulations.

Compared to our previous publication on this topic \cite{burov2024}, concrete major improvements are:
\begin{itemize}
    \item Execution of first‑order Trotterized NMR Hamiltonian evolution for molecules with up to 34 spins on current IBM Heron hardware, with two‑qubit circuit depths up to $250$ after optimization.
    \item Demonstration of an error suppression and mitigation pipeline based on Q-CTRL's Fire Opal tool, achieving up to $22$ times improvement in MSE between hardware and noiseless FID on a 22‑spin system, enabling recovery of some important spectral features.
    \item Cross‑platform comparison of the same NMR simulation on superconducting (IBM Heron) and trapped‑ion (IonQ Forte Enterprise) hardware, including runtime and error‑metric comparisons.
\end{itemize}

\section{Results\label{sec:results}}

\begin{figure*}[ht!]
    \centering
    \includegraphics[width=\textwidth]{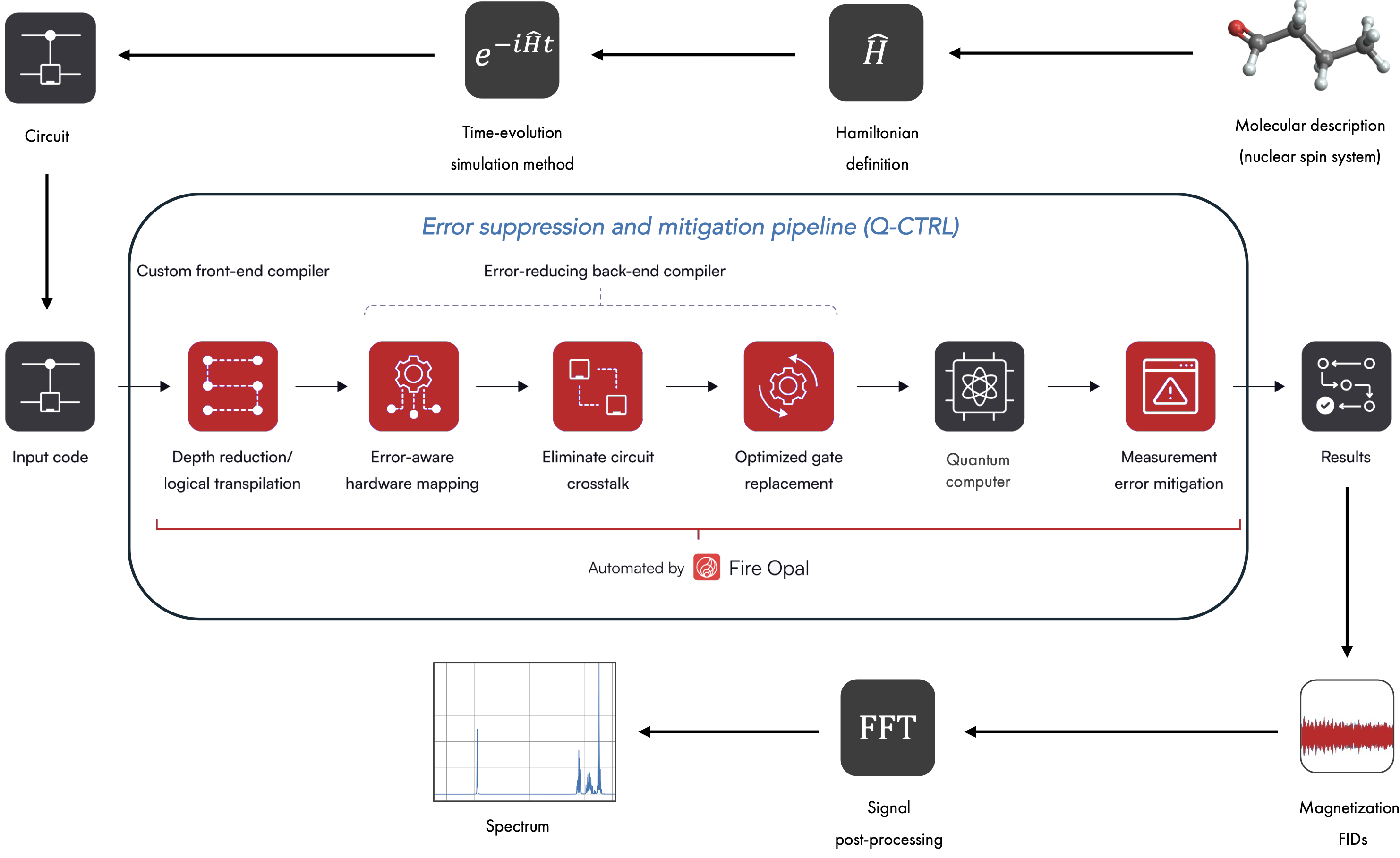}
    \caption{The complete computation pipeline of the paper: from the molecular description (upper right), i.e. the description of the nuclear spin system, we define the Hamiltonian of interest. Its time-evolution is encoded into a quantum circuit depending on the quantum Hamiltonian simulation method chosen, e.g. Trotterization in this work. The error suppression and mitigation pipeline using Q-CTRL Fire Opal tool, fine-tuned to our simulation, is implemented (middle). The result are then collected (e.g. magnetization operators) to build the FIDs. We perform an FFT to get the final NMR 1D spectrum (lower left).}
    \label{fig:pipeline}
\end{figure*}

The NMR Hamiltonian is a Heisenberg Hamiltonian. It is commonly written as
\begin{equation}\label{NMRH1}
     H_{lab}=-\sum^N_{k=1} \gamma_k \vec{I}_k(1-\sigma_k)\vec{B}_0 + 2\pi\sum_{k<l}\vec{I}_kJ_{kl}\vec{I}_l
\end{equation}
for liquid state NMR experiments with spin-$\frac{1}{2}$ nuclei, which is the case considered in this work. Term-by-term it can be understood as: $\gamma_k\vec{I}_k\vec{B}_0$ describes the interaction of spins $\vec{I}_k$ of unshielded nuclei that have gyromagnetic ratios $\gamma_k$ with the constant magnetic field $\vec{B}_0$ conventionally oriented along the $z$-axis, $\vec{B}_0=B_0 \vec{z}$; $-\gamma_k\vec{I}_k\sigma_k\vec{B}_0$ describes the modification $\sigma_k$ to this interaction due to the shielding by the electrons of the molecule, the \emph{chemical shift}; $2\pi\sum_{k<l}\vec{I}_kJ_{kl}\vec{I}_l$ describes the spin-spin interaction mediated by Fermi contact interaction with electrons. In general, both $\sigma_k$ and $J$ can be a tensor, in which case the interaction is anisotropic. In high-field liquid NMR, however, it can often be assumed that the interaction is isotropic so that both the chemical shifts and the $J$-couplings are scalar quantities to a good approximation, which is the regime in our work.

The leading terms in the Hamiltonian come from the parts proportional to $\gamma_k \vec{B}_0$, called the Zeeman part. It is thus customary to work in the rotating frame at the frequency of a reference nucleus species $S$, $\omega_S^{ref} = \gamma_S B_0$. Then, the Hamiltonian reduces to
\begin{equation}\label{NMRH2}
     H_{rot}=\sum^N_{k=1} \omega_k I_k^z + 2\pi\sum_{k<l} J_{kl}\vec{I}_k\vec{I}_l,
\end{equation}
where $\omega_k = \sigma_k\gamma_k B_0$ denotes the chemical-shift–induced offset of spin $k$ in the rotating frame. We simulate the high-field NMR FID in the Heisenberg model without the secular approximation, keeping the full $\vec{I}_k\vec{I}_l = I^X_k I^X_l + I^Y_k I^Y_l + I^Z_k I^Z_l$.

In high‑field liquid‑state NMR the equilibrium state is very close to maximally mixed, with a small polarization. The FID is linear in this small deviation from the identity, so we simulate the dynamics of a pure transverse‑magnetization state representing the deviation density matrix. The identity component is ignored, as it does not contribute to the signal. This corresponds to working directly with the deviation-density operator and modelling it by pure transverse state $\ket{+}^{\otimes{N}}$, up to an overall scale factor, same as in~\cite{burov2024}. In the easiest case in NMR experiments, this state $|\psi_0\rangle$ can be constructed with a $\pi/2$ pulse to rotate the population imbalance from the $z$ axis to be in the transverse plane. For this work, we stay in the zero-temperature limit by evolving a pure state.

We employ only one Trotter step for \emph{every time point} of the time evolution,
\begin{equation}
\label{eq:trotter}
|\psi(t)\rangle = e^{-i\hat{H}_{rot} t}|\psi_0\rangle \approx \prod_j e^{-i\hat{H}_j t}|\psi_0\rangle,
\end{equation}
where $\hat{H}_j$ are the Hamiltonian terms, the interactions of individual spins with the field and the spin-spin interactions. The spin operators are written as $\hat{I}_{x,y,z} = \frac{1}{2}\hat{\sigma}_{x,y,z}$ for spin $\frac{1}{2}$-nuclei, using the Pauli operators, for a direct mapping of spins onto qubits.

We employ the same Hamiltonian decomposition and product formula as in~\cite{burov2024} for the one-Trotter step. Our goal is to primarily investigate deep circuits for NMR spectroscopy simulations and how we can overcome noise on NISQ devices, providing a path for future quantum advantage in NMR spectroscopy simulations, not necessarily demonstrating this advantage already now. Suppose that we employ more Trotter steps or work in the Liouville space with density matrices. In that case, the corresponding circuits will be even deeper than the already deep circuits we investigate, so that the increase in accuracy in our approximation method is negatively balanced by the increase in errors in the circuit. With future, better quantum hardware, our pipeline can be employed with more Trotter steps and working in the Liouville space, to deliver results that can potentially demonstrate quantum advantage.

As shown in previous studies~\cite{burov2024,khedri2024}, overcoming the noise from NISQ quantum hardware is a crucial challenge. We employ an error suppression and mitigation pipeline based on Q-CTRL Fire Opal tool~\cite{fireopal} integrated into our entire computation pipeline presented in Figure~\ref{fig:pipeline}.

We start from the molecular description and use the spin system matrix of the molecule of interest; see the Appendix~\ref{appendix:spinmatrix} for the details of the spin matrices of our spin systems (upper right corner of Figure~\ref{fig:pipeline}). With this matrix, we build the Hamiltonian $H_{rot}$, then generate a parametrized quantum circuit that represents the time evolution of the Hamiltonian with one Trotter step (upper left corner of Figure~\ref{fig:pipeline}). Then, we substitute the time parameter values into this parametrized circuit, one for each time point of the simulation. The resulting circuits represent the input code that is processed by the error suppression and mitigation pipeline (central row of Figure~\ref{fig:pipeline}). Magnetization operators
\begin{equation}
    M_X^{S}=\sum_{j\in S}I^X_j,
    M_Y^{S}=\sum_{j\in S}I^Y_j
\end{equation} in the transverse plane are measured for a given spin species $S$, providing FID signals in the $X$ and $Y$ directions (lower right corner of Figure~\ref{fig:pipeline}) for each time point. The sum is only done on the spins under consideration for the 1D spectrum, and not on all spins of the spin system. We then post-process the signal with padding, using a \textit{fast Fourier transform} (FFT), and rescaling the frequencies to get everything in ppm (part per million), to obtain the final simulated 1D NMR spectrum of the given molecule (lower left corner of Figure~\ref{fig:pipeline}).

\subsection{Circuit depth reduction}

\begin{figure*}[ht!]
    \centering
    \includegraphics[width=\textwidth]{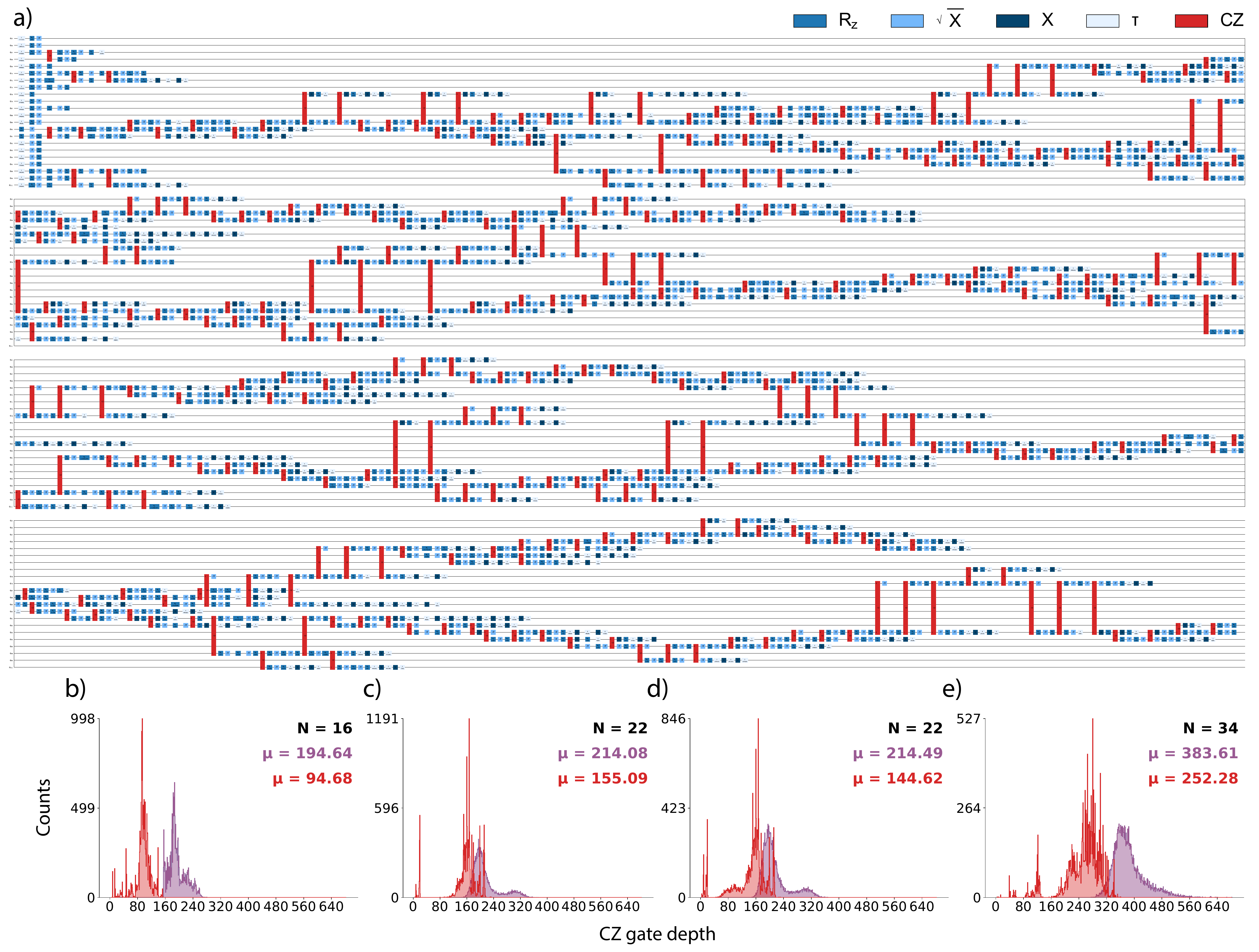}
    \caption{a): example of a transpiled simulation circuit for the $^1$H spectrum of the symmetric P-H molecule (22 spins) for the last obtained time point. b), c), d), e): histograms of frequency distribution of the two-qubit (CZ gate) depths of the quantum simulation circuits per time point (in magenta for the computation on \ibm\ only, in red for the computation on \ibm\ using the error suppression and mitigation pipeline), for difluoroheptane (N=16 spins, fluorine spectrum), two measurements of the symmetric molecule (N=22 spins, c) phosphorus spectrum and d) hydrogen spectrum), and the phosphorous cluster (N=34 spins, phosphorus spectrum), respectively. The depth reduction achieved with the error mitigation pipeline is visible when comparing the mean depth ($\mu$) of each distribution across all four simulations.}
    \label{fig:circuit}
\end{figure*}

In Table~\ref{table:gatecount}, we give the two-qubit circuit depth in all our computations. We compare the IBM runs without the optimized error suppression and mitigation pipeline, the IBM runs including it, as well as the IonQ run of the symm\_P spectrum. Note that for the run on \ionq\ the effective circuit depth corresponds to the number of two-qubit gates of the circuit because each gate is run sequentially on the trapped-ion system. The number of one- and two-qubit gates is the same for all time points on \ionq\ thanks to the all-to-all connectivity of the quantum chip, such that the optimized circuit topology stays the same for all time points. The circuit depth is reduced by a factor of two for the DFH spectra, and by 28\% to 34\% for all other spectra. The reduction is even more significant on the IonQ QPU, from an average of 145 for the optimized IBM run down to 36 (or effectively 69 as all operations on IonQ are sequential) for the optimized IonQ run for the symm\_P spectrum, an additional reduction of a factor of two to four due to the all-to-all-connectivity nature of the IonQ chip. The results on IBM \ibm\ are illustrated in Figure~\ref{fig:circuit} where the distribution
 of two-qubit circuit depths across all time points for each spectrum simulation is presented: b) for the $^{19}$F difluoroheptane spectrum, c) and d) for the $^1$H and $^{31}$P symmetric P-H system spectra, respectively, and e) for the $^{31}$P phosphorous cluster spectrum. In all four cases, the circuit depth distribution when using the error suppression and mitigation pipeline (in red) is peaked on lower values compared to the distribution without the pipeline (in magenta), and the difference between the two distributions is significantly visible for the difluoroheptane and phosphorous cluster spectra. Figure~\ref{fig:circuit} a) displays an example of a transpiled circuit on IBM \ibm\ for the simulation of the last time point in the computation of the symmetric P-H proton spectrum, using the error suppression pipeline.

\begin{table*}[t!]%
  \centering
  %\footnotesize
  \begin{tabular}{ll|cccc}
    & & \# of spins (=qubits)
    & 2-qubit depths
    & \# of 1-qubit gates & \# of 2-qubit gates \tabularnewline
      \hline
      DFH & IBM
    & $16$ & $195\pm 25$ & $1,820\pm 39$ & $450\pm 16$\tabularnewline
    & IBM + Q-CTRL
    & $16$ & $95\pm 25$ & $1,876\pm 541$ & $278\pm 85$\tabularnewline
      symm\_H & IBM
    & $22$ & $214\pm 39$ & $1,575\pm 44$ & $433\pm 16$\tabularnewline
    & IBM + Q-CTRL
    & $22$ & $155\pm 46$ & $2,097\pm 547$ & $312\pm 87$\tabularnewline
      symm\_P & IBM
    & $22$ & $214\pm 39$ & $1,568\pm 42$ & $433\pm 16$\tabularnewline
    & IBM + Q-CTRL
    & $22$ & $145\pm 49$ & $2,020\pm 551$ & $298\pm 88$\tabularnewline
    & IonQ + Q-CTRL
    & $22$ & $36$ & $250$ & $69$\tabularnewline
      phosphorous cluster & IBM
    & $34$ & $384\pm 46$ & $3,037\pm 65$ & $836\pm 29$\tabularnewline
    & IBM + Q-CTRL
    & $34$ & $252\pm 53$ & $3,858\pm 818$ & $615\pm 136$\tabularnewline
    \hline
  \end{tabular}
  \normalsize
  \caption{\label{table:gatecount} The number of qubits, the two-qubit circuit depth, as well as the total number of one- and two-qubit gates for our spectrum simulations as a function of the quantum hardware architecture and the use of the advanced error suppression and mitigation pipeline. We have averaged over all the time points and magnetizations and indicated the $1\sigma$ standard deviation. Note that the effective depth for the IonQ run is actually equal to the amount of two-qubit gates as gate operations are sequential on the trapped-ion quantum computer.}
\end{table*}

We also indicate the total number of one- and two-qubit gates, excluding measurement gates, potential barrier, and time delay. The error suppression and mitigation pipeline always leads to a reduction in the average number of two-qubit gates, which are the most affected by noise. On the other hand, the one-qubit gate count is always larger when using the pipeline compared to the direct output from the \ibm\ system, which is due to dynamical decoupling which increases the number of one-qubit gates in order to increase the fidelity of the circuit. The stronger variation in gate count from one circuit to the other is also reflected by the increase in the $1\sigma$ standard deviation when comparing the bare output from the QPU and the gate count for the optimized circuits. The circuit optimization is done per circuit, which implies that the pipeline will find e.g. different layout mapping and transpilation from one run to the other, use different dynamical decoupling, etc. All of this has an impact on the final optimized circuit and on the gate count, explaining the pronounced increase of standard deviation.

\subsection{Error reduction on the FID signal}

\begin{figure*}[ht!]
    \centering
    \subfloat[First 800 time points]{
    \includegraphics[scale=0.49]{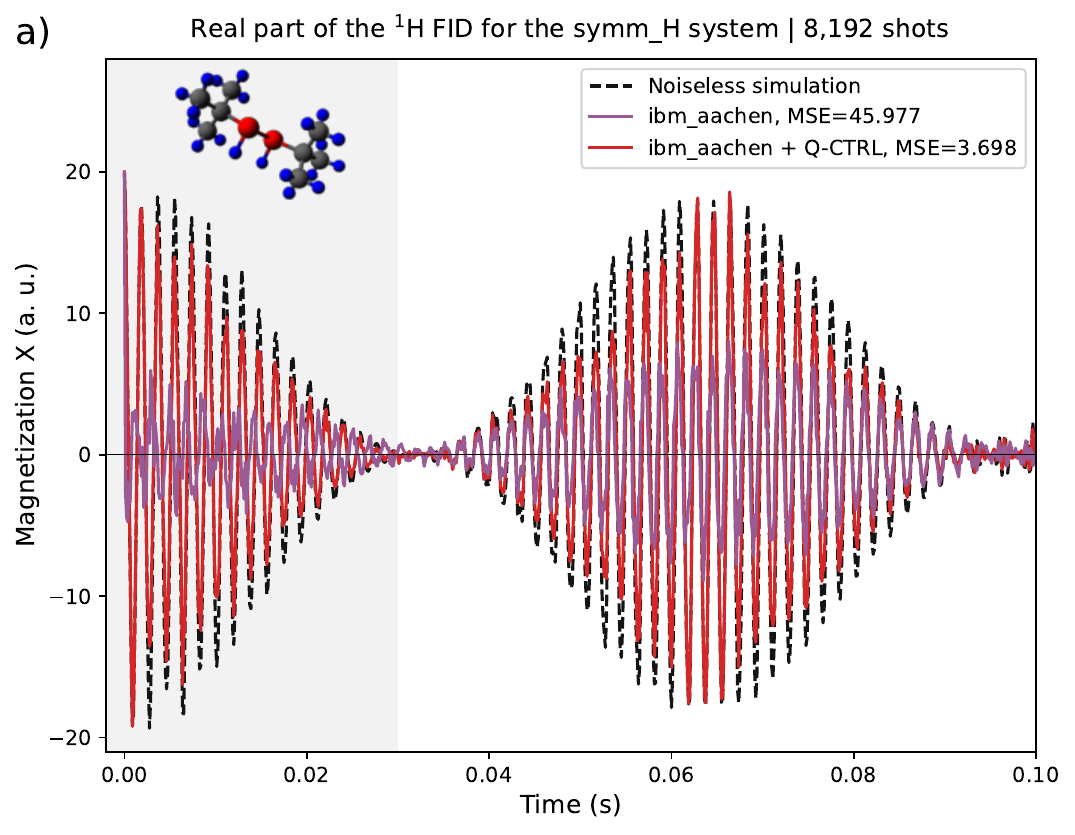}
    \label{fig:FID:main}}
    \subfloat[Zoom on the first 240 time points]{
    \includegraphics[scale=0.49]{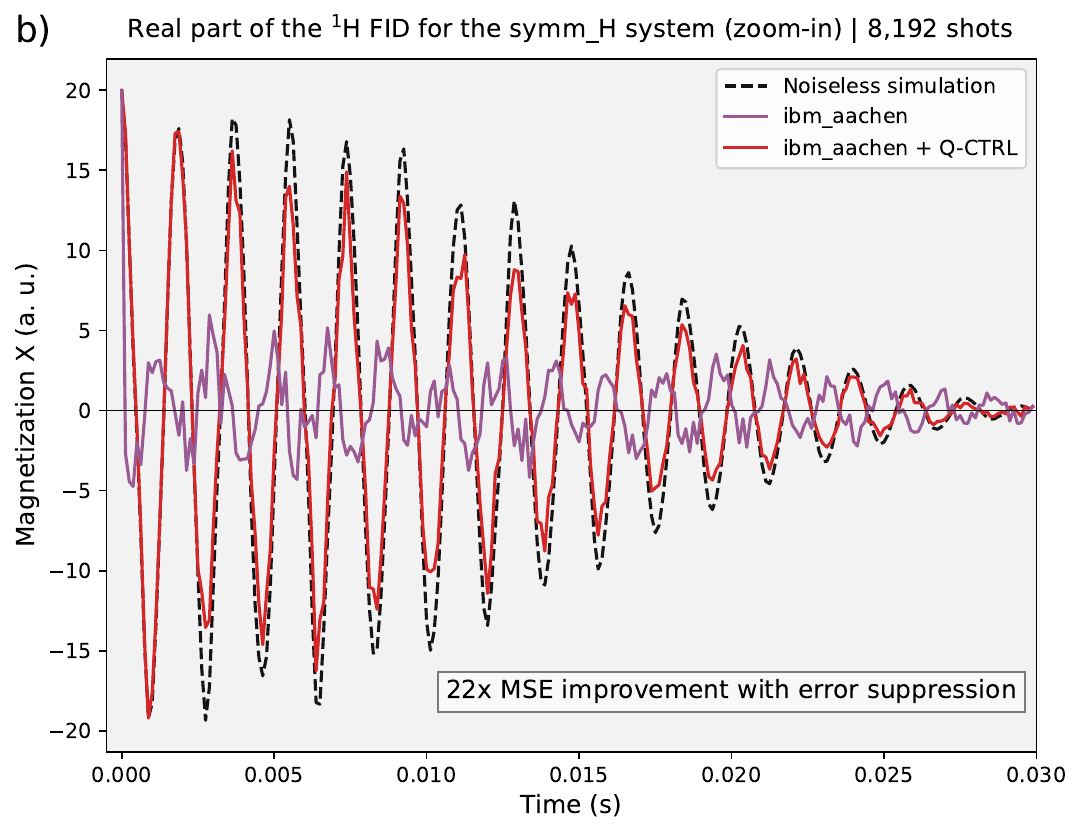}
    \label{fig:FID:zoom}}
    \caption{First 800 time points (left) and zoom (gray area) on the first 240 time points (right) of the real part of the FID ($X$ magnetization) for the symm\_H system as a function of time (in seconds). We compare the computation on \ibm\ quantum computer (in solid magenta), and the simulation on \ibm\ together with the error suppression pipeline (in solid red) against the computation on an ideal noiseless quantum simulator (in dashed black). On the 800 time points, the uncorrected quantum computation has $\text{MSE}=45.977$ and a cosine similarity of 0.511 compared to the noiseless simulation, while the computation using the Q-CTRL error suppression pipeline has $\text{MSE}=3.698$ and a cosine similarity of 0.985. All computations have been performed with $8,192$ shots.}
    \label{fig:FID}
\end{figure*}

To illustrate the impact of the error suppression and mitigation pipeline, we display in Figure~\ref{fig:FID} the first few times points of the $^1$H FID for the symmetric P-H molecule (symm\_H), focusing on the real part of the FID (coming from the $X$ magnetization). Each time point has been generated with runs using $8,192$ shots. The figures show a comparison between the ideal noiseless computation as performed on the local qiskit {\tt aer} quantum simulator (dashed black lines), the computation on \ibm\ quantum computer without any error suppression or error mitigation (solid magenta lines), and the computation on \ibm\ using our entire pipeline with Q-CTRL error suppression and mitigation (solid red lines). In Figure~\ref{fig:FID:main} we display the first 800 time points, while Figure~\ref{fig:FID:zoom} represents a zoom-in on the first 250 time points. It is immediately clear that the impact of our pipeline is very important, especially on the zoom-in figure: the magenta line barely captures the actual dynamics represented by the dashed black line, while the red line, representing the outcome of our full pipeline, nearly recovers the entire noiseless, ideal simulation. 

The \textit{mean square error} (MSE) between the computation on the quantum computer and the noiseless simulation, calculated over the 800 first time points, amounts to 45.977 for the bare quantum computation, while the MSE for the computation using the full pipeline is only 3.698. This is a factor of 12 improvement in the quality of the computation. The same improvement is quantitatively obtained when comparing the cosine similarity of the noiseless and hardware-generated signals $\vec{S_{nl}}$ and $\vec{S_{hw}}$
\begin{equation}
    C=\frac{\vec{S_{nl}}\cdot\vec{S_{hw}}}{\norm{\vec{S_{nl}}}\cdot\norm{\vec{S_{hw}}}}.
\end{equation}
The value $C=0.51$ on the bare computation improves up to $C=0.99$ when using the error suppression and mitigation pipeline. The MSE improvement even reaches a factor of 22 when comparing the distributions on the first 240 time points, as shown in Figure~\ref{fig:FID:zoom}. The accuracy of the quantum computation has thus tremendously increased when using the full error suppression and mitigation pipeline.

\subsection{Results for the 1D spectra}

\begin{figure*}[ht!]
    \centering
    \includegraphics[width=\textwidth]{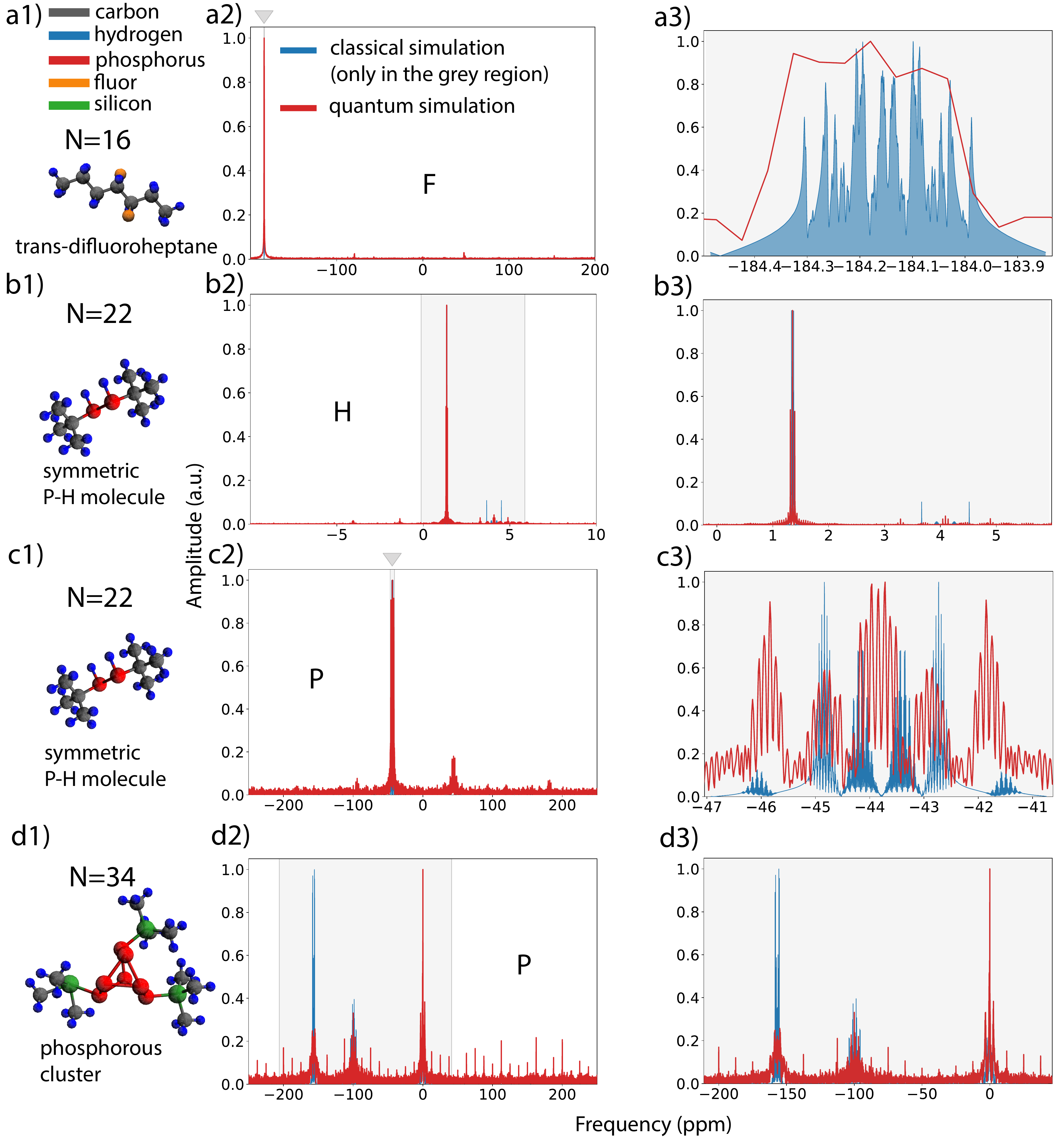}
    \caption{a) Difluoroheptane, b) and c) symmetric P-H molecule, d) phosphorous cluster molecules (column 1) and 1D NMR spectra (columns 2 and 3) with the quantum result on IBM \ibm\ quantum computer using the error suppression and mitigation pipeline. Column 3) shows the zoom-ins on the regions of interest that are also simulated in SPINACH (in blue). The observed nuclei are Fluorine for a), Hydrogen for b), and Phosphorus for c) and d).}
    \label{fig:spectra}
\end{figure*}

We present our final results after processing the FIDs we have obtained from the quantum computation. We have applied a zero padding of the FID signals, and then an FFT to generate the 1D NMR spectra. We have simulated the four hard experiments from the SPINACH example folder~\cite{spinach} and display the results in Figure~\ref{fig:spectra} on IBM \ibm\ using our full error suppression and mitigation pipeline. All spectra have been simulated for the frequency range that covers the usual ppm range observed in the measured nuclei with $8,192$ time points. The results presented in the rows a), b), c), and d) are associated with the difluoroheptane (DFH) $^{16}$F 1D spectrum (with $4,096$ shots), the symmetric P-H $1^H$ 1D spectrum (symm\_H, with $8,192$ shots), the symmetric P-H $^{31}$P 1D spectrum (symm\_P, with $4,096$ shots), and the phosphorous cluster $^{31}$P 1D spectrum (with $8,192$ shots), respectively. The molecules are displayed in column 1, while the full spectra are displayed in column 2 and the zoom-in on the interesting region is presented in column 3. The red lines represent the quantum results, while the blue lines represent the restricted classical simulation using SPINACH. We stress that the classical simulation is only performed in the region of interest (the gray region). Given that the number of main peaks and their localization are related to the difference in chemical shifts, it is not expected to witness peaks outside of the gray regions. The classical simulation for the four cases we present being quite computationally intensive, it is wiser to focus only on the region of interest for the classical simulation with SPINACH.

Compared with the results in~\cite{burov2024}, the impact of the quantum noise is visibly reduced, in particular for the DFH and symmetric P-H molecules. The salient features of all of the spectra are reproduced in the quantum computations, even for the hardest simulation of the phosphorous cluster: the position of the main peaks is correct. In Figure~\ref{fig:spectra} a3) we see the indication of the substructure of the peak as well, but the resolution is not good enough to be more precise. In Figure~\ref{fig:spectra} d3), for the phosphorous cluster, indication of the substructure is also present with a good alignment between SPINACH and the quantum computation. This is the first time this has been achieved for such a large spin cluster. Figure~\ref{fig:spectra} c3) also displays the zoom-in for the symm\_P spectrum. There is a distortion of the subpeaks that is not aligned with the SPINACH reference computation.

\begin{figure*}[ht!]
    \centering
    \includegraphics[width=\textwidth]{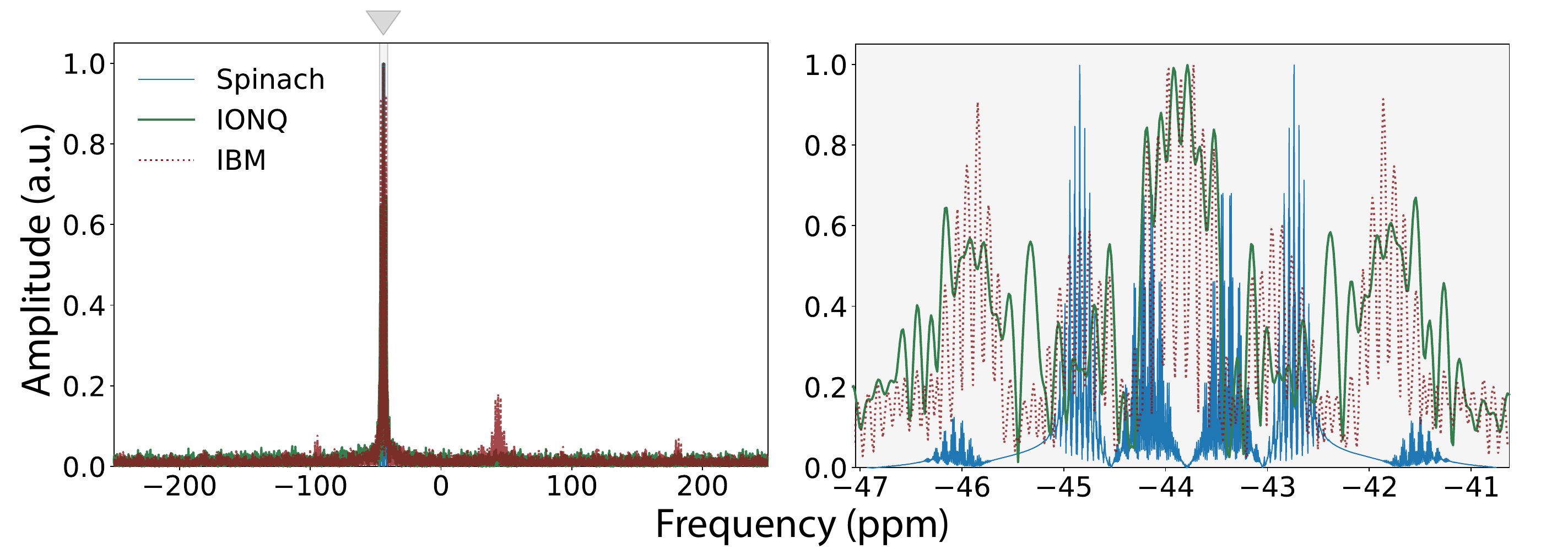}
    \caption{Spectra of symm\_P simulated on IonQ \ionq\ (solid, green line) and on IBM \ibm\ (dotted, red line).}
    \label{fig:spectrum_ionq}
\end{figure*}

We have also performed the computation of the symm\_P spectrum on the IonQ \ionq\ quantum computer, using $4,096$ time points, each of them calculated with two runs with 700 shots. In Figure~\ref{fig:spectrum_ionq}, we present the results comparing the IBM results (dotted, red lines) and the IonQ results (solid, green lines) together against the classical computation with SPINACH (solid blue lines). The two quantum computations yield similar results in the region of interest (in gray), with the IonQ computation slightly more aligned with the classical reference result (right-hand side plot in Figure~\ref{fig:spectrum_ionq}) but still not precise enough to capture the complete peak substructure. However, for the full ppm range, as depicted in Figure~\ref{fig:spectrum_ionq} (on the left-hand side), the IonQ result does not show spurious peaks in the positive ppm region contrary to the IBM computation, indicating a better simulation with less impact from quantum noise. Note that the IonQ and IBM datasets are not matched in time-point/shot budgets, so this qualitative comparison should be interpreted cautiously.

\subsection{Computation time}

The \ibm\ quantum computer can perform much faster operations, thanks to the superconducting-qubit technology, as shown in Table~\ref{table:specifications}. Each run on \ibm\ contains between 30 and 80 circuits submitted to the system (as primitive unified blocks), allowing for a low classical latency between each time-point run. For the symm\_P spectrum, the total QPU runtime on \ibm\ amounts to 5~hours, 20~minutes, and 10~s with an average of $1.18\pm 0.03$~s per circuit run with $4,096$~shots. The equivalent computation on the IonQ \ionq\ quantum computer amounts to a total runtime of 105~hrs, 58~min, and 55~s to generate all the time points, corresponding to an average of $47\pm 2$~s for one run with 700 shots. We thus need 40x less time on the IBM quantum computer than on the IonQ quantum computer to collect all of our data. We present in Table~\ref{table:runtime} the runtime statistics for all spectrum simulations we have performed. Note that the time needed for the readout error mitigation is not accounted for in the numbers presented. It amounts to 40 to 50~s on \ibm\ (for each spin system) and around 1~min and 30~s on \ionq\ (for the symm\_P spin system), for all time points at once in both cases.

\begin{table}[ht!]%
  \centering
  %\footnotesize
  \begin{tabular}{l|cc}
    & QPU runtime
    & QPU runtime/circuit \tabularnewline
      \hline
      DFH (IBM) & 5~hrs, 15~min, 17~s & $1.16\pm 0.04$~s \tabularnewline
      symm\_P (IBM) & 5~hrs, 20~min, 10~s & $1.18\pm 0.03$~s \tabularnewline
      symm\_H (IBM) & 10~hrs, 40~min, 37~s & $2.35\pm 0.05$~s \tabularnewline
    phosphorous & 11~hrs, 3~min, 9~s & $2.4\pm 0.2$~s \tabularnewline
    cluster (IBM) &  & \tabularnewline
      symm\_P (IonQ) & 105~hrs, 58~min, 55~s & $47\pm 2$~s \tabularnewline
    \hline
  \end{tabular}
  \normalsize
  \caption{\label{table:runtime} Total QPU runtime on IBM \ibm\ quantum computer (first four rows) and on IonQ \ionq\ quantum computer (last row) for the FID computations of the four spin systems, as well as the average QPU runtime per circuit run. The DFH and symm\_P simulations (on \ibm) have been performed with $4,096$ shots while the symm\_H and phosphorus cluster computations have been performed with $8,192$ shots. The symm\_P simulation on \ionq\ has been performed with 700 shots.}
\end{table}

\section{Discussion and conclusion\label{sec:discussion}}

The path to quantum utility and eventually quantum advantage necessitates the execution of deep circuits with significant two-qubit depth. As NMR is a promising case to witness such an advantage, it is crucial to create a working pipeline enabling deep circuit execution for NMR quantum Hamiltonian simulation on NISQ devices. We have, for the first time, successfully demonstrated that with one Trotter step per time point, combined with an advanced error suppression and mitigation pipeline based on Q-CTRL Fire Opal tool, it is possible to execute circuits up to a depth of 250 with a significant reduction of the impact of quantum noise, quantified by the up to 22x reduction in the mean-square error when comparing noiseless simulation with the computation on the quantum hardware. We have simulated NMR 1D spectra for four spin systems identified as classically hard, up to 34 spins with the phosphorous cluster, slightly beyond the practical Liouville limit $N=32$ identified in previous work~\cite{elenewski2025} for unrestricted Liouville propagation. Results both on IBM \ibm\ and IonQ \ionq\ quantum hardware have been presented, the IonQ results pointing to a slight reduction of the impact of quantum noise with the avoidance of unwanted artifacts in the 1D spectrum. It should be noted that the environment in liquid NMR could impact the NMR dynamics, with the interaction of the spin system with the atomic structure of the bath. This type of open-system dynamics remains largely unexplored in NMR simulations and is beyond the scope of this study, but constitutes an interesting avenue for future work as the quantum approach could in principle include these many additional external spins or use quantum thermodynamics concepts, while the classical approach would need to rely on strong approximation to handle the complexity of the problem.

We have compared our quantum computation with a classical benchmark provided by SPINACH. Although the position of the main peaks is successfully simulated, and the overall substructure of the peak is present, especially for the phosphorous cluster, there is still work to be done to get high-quality results capturing all the details of the peaks substructure. Such inaccuracies are expected with the one Trotter step approximation used here for the purpose of exploring the reduction of noise in deep circuits. Better quantum hardware will allow for deeper circuit executions, in turn opening up the possibility to go beyond the simple one Trotter step approximation. Furthermore, there exist several other alternatives for simulating Hamiltonian dynamics which would help getting more accurate results without inflating the depth of the circuits. Qubitization~\cite{Low2019hamiltonian}, using Taylor-series expension~\cite{Berry2015} or Dyson-series expension~\cite{Kieferov2019} has been proven to be very efficient, with optimal scaling both in time $t$ for the time-independent Hamiltonian (such as the one used in NMR simulations) and inverse error $1/\varepsilon$ in an additive way. Other alternatives have also emerged, promising even the ability to have no discretization error at all~\cite{Granet2024}, or the spin echo technique, which reduces the circuit depth quite dramatically~\cite{Abanin2025,Zhang2025}. Novel quantum techniques developed for the computation of molecular free energies also provide interesting avenues for Hamiltonian time evolution~\cite{huang2025}. These alternative paths have not been explored yet for NMR quantum simulation, and combined with an advanced error suppression and mitigation pipeline as demonstrated here for the first time in NMR quantum simulations, they may provide the first compelling results of quantum utility in NMR spectroscopy.

\section{Material and Methods\label{sec:methods}}

\subsection{Setup for the classical computation with SPINACH}

\begin{table*}[t!]%
\centering
  \begin{tabular}{l|ccccc}
    & $^{19}$F DFH
    & $^{31}$P symm\_P (IBM)
    & $^{31}$P symm\_P (IonQ)
    & $^1$H symm\_H
    & $^{31}$P phosphorous cluster\tabularnewline
      \hline
      \# of qubits
    & $156$ & $156$ & $36$ & $156$ & $156$\tabularnewline
      Coherence time $T_1$ ($\mu$s)
    & $210$ & $213\pm 2$ & $1.88\times 10^8$ & $210\pm 7$ & $210\pm 2$\tabularnewline
      Coherence time $T_2$ ($\mu$s)
    & $188$ & $179\pm 8$ & $9.5\times 10^5$ & $189\pm 1$ & $177\pm 4$\tabularnewline
      One-qubit gate time ($\mu$s)
    & $0.032$ & $0.032$ & $63$ & $0.032$ & $0.032$\tabularnewline
      Two-qubit gate time ($\mu$s)
    & $0.068$ & $0.068$ & $650$ & $0.068$ & $0.068$\tabularnewline
      Readout time ($\mu$s)
    & $2.60$ & $2.60$ & $250$ & $2.60$ & $2.60$\tabularnewline
      One-qubit gate error rate
    & $2.07\times 10^{-4}$ & $2.08\times 10^{-4}$ & $(2.1\pm 0.3)\times 10^{-4}$ & $2.21\times 10^{-4}$ & $(2.25\pm 0.06)\times 10^{-4}$\tabularnewline
      Two-qubit gate error rate
    & $1.93\times 10^{-3}$ & $(1.91\pm 0.02)\times 10^{-3}$ & ($7.66\pm 1.34)\times 10^{-3}$ & $(2.05\pm 0.05)\times 10^{-3}$ & $(2.17\pm 0.08)\times 10^{-3}$\tabularnewline
      Readout error rate
    & $7.81\times 10^{-3}$ & $(7.63\pm 0.13)\times 10^{-3}$ & $(6.08\pm 0.63)\times 10^{-3}$ & $(7.57\pm 0.46)\times 10^{-3}$ & $(8.00\pm 0.05)\times 10^{-3}$\tabularnewline
    \hline
  \end{tabular}%
  \caption{\label{table:specifications} Salient parameters from the technical specifications of the IBM (\ibm) and IonQ (\ionq) devices, depending on the NMR spectrum simulation, at the time of the quantum computations on the respective devices. The values quoted here are the average of the corresponding median values (over all qubits) of the quantum devices, taken over both magnetizations and all the time points. When significant, the standard deviation is displayed as well. Note that the two-qubit gate on \ibm\ is the CZ gate while it is the M{\o}lmer-S{\o}rensen gate on \ionq. The readout on \ionq\ is done on all qubits at once.}
\end{table*}

As a benchmark for our pipeline, we have used four spin systems: DFH, symm\_H, symm\_P and phosphorous cluster. The simulations were designed to correspond to the simulations of these spin systems in the examples' folder of the SPINACH~\cite{spinach} library. To provide some details on those classical simulations, as an example, the hardest of the presented molecules requires a GPU to simulate with SPINACH and several hours of compute time. This simulation is performed in the Liouville space with spherical-tensor formalism in a restricted state space; only the basis states up to global coherence order 1 are kept (longitudinal and single-quantum coherences). Additionally, for protons, only longitudinal states are kept, and three methyl‑like proton triplets are declared to have permutation symmetry and are tracked as fully symmetric irreducible representations of those symmetry groups.

\subsection{Technical set-up on the quantum computers}

We have performed our simulations on two different quantum hardware modalities. For the DFH ($^{19}\text{F}$ spectrum), proton spectrum of the symmetric molecule, and phosphorous cluster ($^{31}\text{P}$ spectra), we have used the 156-qubit IBM superconducting-qubit quantum system \ibm\ located in Germany, based on the Heron chip with a heavy-hex lattice topology, while the $^{31}\text{P}$ spectrum of symmetric molecule has been simulated both on \ibm\ and on the 36-qubit trapped-ion quantum system \ionq\ from IonQ (located in Switzerland on the uptownBasel campus hosting QuantumBasel), which has an all-to-all connectivity of the qubits on the chip and uses $^{2}S_{1/2}$ states of $^{171}\text{Yb}^{+}$ ions as the qubit states.

Of the 36 qubits of the \ionq\ system, we have used 22 qubits for the simulation of the symm\_P spectrum (one qubit per spin in the molecule) and used the measurement on the last two qubits, corresponding to the $P$ atoms, for the computation of the FID. We have used 16, 22, 22, and 34 qubits on \ibm\ for the simulation of the DFH, symm\_H, symm\_P, and the phosphorous cluster, respectively. For the DFH spectrum the measurement has been performed on qubits 4 and 14 corresponding to the $F$ atoms; for the symm\_P spectrum it is the same protocol as on the \ionq\ system, while for the symm\_H spectrum the measurement has been performed on the first 20 qubits corresponding to the $H$ atoms; for the phosphorous cluster the measurement has been performed on the last 7 qubits corresponding to the $P$ atoms.

We have generated the time series data for all runs on the \ibm\ system with $8,192$ points, each of them generated with two runs of $8,192$ shots each to collect the $M_X$ and $M_Y$ magnetizations ($4,096$ shots for DFH and symm P, where we empirically found that $4,096$ shots provided sufficient signal‑to‑noise for the two measured spins, in contrast to the $8,192$ shots used when averaging over many spins). The time series data for the symm\_P spectrum on \ionq\ has been generated with $4,096$ time points, each of them with two runs of $700$ shots each. This choice for the number of shots on the IonQ quantum computer follows the protocol detailed in the Appendix~\ref{appendix:ionqprotocol} to balance runtime with accuracy. To mitigate the drift of system specifications during our computation time window on the IonQ quantum system, we have generated the time points in a random order and not sequentially, to average out this drift. 

\subsection{Error suppression and mitigation pipeline}
\label{sec:methods:qctrl}

We present in this subsection the error suppression and mitigation pipeline we have used to generate our results. We have used a tuned version of Fire Opal~\cite{fireopal}, the automatic error suppression and mitigation tool developed by Q-CTRL~\cite{Mundada2023}. As presented in Figure~\ref{fig:pipeline} this is a central part of our computation pipeline and it relies generically on five main steps. An exact implementation slightly varies from one hardware modality to the other with potential adaptation or a subset selection of these steps.

\paragraph{Logical transpilation}

The first step is the transpilation of the logical circuit to the physical device, ensuring an optimal mapping to the native gate of the target hardware. There is a sequence of compiler passes to reduce already at this stage the circuit depth thanks to mathematical combinations (for example, combining several gates into one single gate whenever possible). This step will already produce different circuits on IBM and IonQ quantum architectures as the set of native gates differs: for example \ibm\ quantum chip uses the CZ gate as the native two-qubit gate while \ionq\ uses the M{\o}lmer-S{\o}rensen gate.

\paragraph{Error-aware hardware mapping}

This step selects the optimal layout of the circuit onto the quantum chip. This is especially important for quantum modalities without all-to-all connectivity such as the IBM system, but it is not restricted to this case. As we do not use all the qubits available on IonQ \ionq, it is also important to select the best subset of qubits relative to their individual specs.

\paragraph{Dynamical decoupling}

The goal of this step is to significantly reduce dephasing and circuit crosstalk (such as $ZZ$ crosstalk), by introducing single-qubit pulse sequences and appropriate delays, using the spin echo mechanism, to cancel out these sources of noise. The underlying pulse sequences are optimized for the target quantum hardware, see for example~\cite{pokharel2018} or~\cite{sutherland2024} for protocols with superconducting qubits and trapped-ion qubits, respectively. Fire Opal uses the hardware characterization and topology to create a new circuit on which an optimal dynamical decoupling protocol is applied, taking care of the dephasing ($T_2$ error) and of the $ZZ$ error. These two sources of error are minimized as much as possible while keeping the number of inserted $X$ gates as low as possible, in a trade-off between their number and the amount of $X$ gate error.

\paragraph{Optimized gate replacement}

Since the original optimizer detailed in \cite{Mundada2023}, the method has been refined to a resynthesis of two-qubit gates. They are first consolidated in two-qubit blocks and then optimally synthesized into native gates such that the total number of two-qubit gates is minimized. In a subsequent step, the same procedure is applied to one-qubit gates as well.

\paragraph{Classical post-processing including measurement error mitigation}

After running the error-suppressed circuit on the target quantum device, the results are post-processed to ensure a mitigation of the measurement errors. The custom error mitigation protocol from Fire Opal scales sublinearly with the number of qubits~\cite{Mundada2023} and uses a technique similar to the M3 technique~\cite{bravyi2021,nation2021} introduced by IBM to reduce the memory overhead of storing the entire measurement confusion matrix. The quantum device is divided into smaller groups of qubits which are mitigated, these groups being automatically chosen to minimize intergroup correlations. The confusion matrix is then calculated and combined with the hardware output bit strings to obtain the final output distributions where measurement errors are mitigated. There are other interesting approaches for error mitigation that would be worth trying as well~\cite{alistair2021,funcke2022,pokharel2024}, including a more complete error mitigation protocol~\cite{aharonov2025importance,aharonov2025reliable}.

On top of the regular Fire Opal tool, a final (problem-agnostic) post-processing step is added. Low-confidence bit strings are identified in the output distribution and smoothly rescaled to reduce their impact on the final expectation values.

\section{Data availability}\label{Data}

The datasets generated and analyzed during this study are available at \href{https://doi.org/10.5281/zenodo.17936955}{10.5281/zenodo.17936955}.

\section{Acknowledgments}\label{Acknowledgment}

We acknowledge the use of IBM Quantum services for this work. The views expressed are those of the authors and do not reflect the official policy or position of IBM or the IBM Quantum team. We thank both QuantumBasel platform team and Q-CTRL technical team for the support and access to the quantum computers, in particular Nuiok Dicaire for her support with the IBM runs and Claire L. Edmunds for her support with the IonQ runs. We acknowledge Ilya Kuprov, Klaus Mayer and Frederik Flöther for useful discussions and feedback. 

%\newpage

\bibliography{bibliography}

@article{demler1,
   title={{Quantum approximate Bayesian computation for NMR model inference}},
   volume={2},
   ISSN={2522-5839},
   url={http://dx.doi.org/10.1038/s42256-020-0198-x},
   DOI={10.1038/s42256-020-0198-x},
   number={7},
   journal={Nature Machine Intelligence},
   author={Sels, Dries and Dashti, Hesam and Mora, Samia and Demler, Olga and Demler, Eugene},
   year={2020},
   pages={396–402}
}

@article{demler2,
   title={{Digital quantum simulation of NMR experiments}},
   volume={9},
   ISSN={2375-2548},
   url={http://dx.doi.org/10.1126/sciadv.adh2594},
   DOI={10.1126/sciadv.adh2594},
   number={46},
   journal={Science Advances},
   author={Seetharam, Kushal and Biswas, Debopriyo and Noel, Crystal and Risinger, Andrew and Zhu, Daiwei and Katz, Or and Chattopadhyay, Sambuddha and Cetina, Marko and Monroe, Christopher and Demler, Eugene and Sels, Dries},
   year={2023},
   pages={eadh2594}
}

@article{google_obrian,
   title={{Quantum Computation of Molecular Structure Using Data from Challenging-To-Classically-Simulate Nuclear Magnetic Resonance Experiments}},
   volume={3},
   issue = {3},
   numpages = {25},
   ISSN={2691-3399},
   url={http://dx.doi.org/10.1103/PRXQuantum.3.030345},
   DOI={10.1103/prxquantum.3.030345},
   number={3},
   journal={PRX Quantum},
   publisher={American Physical Society (APS)},
   author={O’Brien, Thomas E. and Ioffe, Lev B. and Su, Yuan and Fushman, David and Neven, Hartmut and Babbush, Ryan and Smelyanskiy, Vadim},
   year={2022},
   pages={030345}
}

@article{EDWARDS2014107,
    title = {{Quantum mechanical NMR simulation algorithm for protein-size spin systems}},
    journal = {Journal of Magnetic Resonance},
    volume = {243},
    pages = {107-113},
    year = {2014},
    issn = {1090-7807},
    doi = {https://doi.org/10.1016/j.jmr.2014.04.002},
    url = {https://www.sciencedirect.com/science/article/pii/S1090780714001086},
    author = {Luke J. Edwards and D.V. Savostyanov and Z.T. Welderufael and Donghan Lee and Ilya Kuprov}
}

@incollection{AGGARWAL2022237,
title = {{Chapter 16 - Advances in liquid-state NMR spectroscopy to study the structure, function, and dynamics of biomacromolecules}},
editor = {Timir Tripathi and Vikash Kumar Dubey},
booktitle = {{Advances in Protein Molecular and Structural Biology Methods}},
publisher = {Academic Press},
pages = {237-266},
year = {2022},
isbn = {978-0-323-90264-9},
doi = {https://doi.org/10.1016/B978-0-323-90264-9.00016-7},
url = {https://www.sciencedirect.com/science/article/pii/B9780323902649000167},
author = {Priyanka Aggarwal and Pooja Kumari and Neel Sarovar Bhavesh}
}

@article{ZHENG20241,
title = {{Advanced solid-state NMR spectroscopy and its applications in zeolite chemistry}},
journal = {Progress in Nuclear Magnetic Resonance Spectroscopy},
volume = {140-141},
pages = {1-41},
year = {2024},
issn = {0079-6565},
doi = {https://doi.org/10.1016/j.pnmrs.2023.11.001},
url = {https://www.sciencedirect.com/science/article/pii/S0079656523000237},
author = {Mingji Zheng and Yueying Chu and Qiang Wang and Yongxiang Wang and Jun Xu and Feng Deng}
}

@article{LI2021116152,
title = {{Solid-state NMR spectroscopy in pharmaceutical sciences}},
journal = {TrAC Trends in Analytical Chemistry},
volume = {135},
pages = {116152},
year = {2021},
issn = {0165-9936},
doi = {https://doi.org/10.1016/j.trac.2020.116152},
url = {https://www.sciencedirect.com/science/article/pii/S0165993620303812},
author = {Mingyue Li and Wei Xu and Yongchao Su}
}

@article{spinach,
    title = {{Spinach – A software library for simulation of spin dynamics in large spin systems}},
    journal = {Journal of Magnetic Resonance},
    volume = {208},
    number = {2},
    pages = {179-194},
    year = {2011},
    issn = {1090-7807},
    doi = {https://doi.org/10.1016/j.jmr.2010.11.008},
    url = {https://www.sciencedirect.com/science/article/pii/S1090780710003575},
    author = {H.J. Hogben and M. Krzystyniak and G.T.P. Charnock and P.J. Hore and Ilya Kuprov}
}

@misc{burov2024,
      title={{Towards quantum utility for NMR quantum simulation on a NISQ computer}}, 
      author={Artemiy Burov and Oliver Nagl and Clément Javerzac-Galy},
      year={2024},
      eprint={2404.17548},
      archivePrefix={arXiv},
      primaryClass={quant-ph},
      url={https://arxiv.org/abs/2404.17548}, 
}

@misc{khedri2024,
      title={{The impact of noise on the simulation of NMR spectroscopy on NISQ devices}}, 
      author={Andisheh Khedri and Pascal Stadler and Kirsten Bark and Matteo Lodi and Rolando Reiner and Nicolas Vogt and Michael Marthaler and Juha Leppäkangas},
      year={2024},
      eprint={2404.18903},
      archivePrefix={arXiv},
      primaryClass={quant-ph},
      url={https://arxiv.org/abs/2404.18903}, 
}

@misc{fratus2025,
      title={{Can a Quantum Computer Simulate Nuclear Magnetic Resonance Spectra Better than a Classical One?}}, 
      author={Keith R. Fratus and Nicklas Enenkel and Sebastian Zanker and Jan-Michael Reiner and Michael Marthaler and Peter Schmitteckert},
      year={2025},
      eprint={2508.06448},
      archivePrefix={arXiv},
      primaryClass={quant-ph},
      url={https://arxiv.org/abs/2508.06448}, 
}

@article{trotter1959product,
  title={On the product of semi-groups of operators},
  author={Trotter, Hale F},
  journal={Proceedings of the American Mathematical Society},
  volume={10},
  number={4},
  pages={545--551},
  year={1959},
  publisher={JSTOR},
  url={https://doi.org/10.2307/2033649},
  DOI={10.2307/2033649}
}

@article{childs2021,
  title = {{Theory of Trotter Error with Commutator Scaling}},
  author = {Childs, Andrew M. and Su, Yuan and Tran, Minh C. and Wiebe, Nathan and Zhu, Shuchen},
  journal = {Phys. Rev. X},
  volume = {11},
  issue = {1},
  pages = {011020},
  numpages = {49},
  year = {2021},
  publisher = {American Physical Society},
  doi = {10.1103/PhysRevX.11.011020},
  url = {https://link.aps.org/doi/10.1103/PhysRevX.11.011020}
}

@misc{fireopal,
  author = {Q-CTRL},
  title = {Fire {O}pal},
  year = {2025},
  howpublished = {https://q-ctrl.com/fire-opal},
  note = {[Online]}
}

@article{Mundada2023,
  title = {Experimental Benchmarking of an Automated Deterministic Error-Suppression Workflow for Quantum Algorithms},
  author = {Mundada, Pranav S. and Barbosa, Aaron and Maity, Smarak and Wang, Yulun and Merkh, Thomas and Stace, T.M. and Nielson, Felicity and Carvalho, Andre R.R. and Hush, Michael and Biercuk, Michael J. and Baum, Yuval},
  journal = {Phys. Rev. Appl.},
  volume = {20},
  issue = {2},
  pages = {024034},
  numpages = {20},
  year = {2023},
  publisher = {American Physical Society},
  doi = {10.1103/PhysRevApplied.20.024034},
  url = {https://link.aps.org/doi/10.1103/PhysRevApplied.20.024034}
}

@misc{elenewski2025,
      title={{Prospects for NMR Spectral Prediction on Fault-Tolerant Quantum Computers}}, 
      author={Justin E. Elenewski and Christina M. Camara and Amir Kalev},
      year={2025},
      eprint={2406.09340},
      archivePrefix={arXiv},
      primaryClass={quant-ph},
      url={https://arxiv.org/abs/2406.09340}, 
}

@article{das2025,
author = {Das, Susanta and Merz, Kenneth M. Jr.},
title = {{Exploring the Frontiers of Computational NMR: Methods, Applications, and Challenges}},
journal = {Chemical Reviews},
volume = {125},
number = {19},
pages = {9256-9295},
year = {2025},
doi = {10.1021/acs.chemrev.5c00259},
URL = {https://doi.org/10.1021/acs.chemrev.5c00259},
}

@article{Low2019hamiltonian,
  doi = {10.22331/q-2019-07-12-163},
  url = {https://doi.org/10.22331/q-2019-07-12-163},
  title = {Hamiltonian {S}imulation by {Q}ubitization},
  author = {Low, Guang Hao and Chuang, Isaac L.},
  journal = {{Quantum}},
  issn = {2521-327X},
  publisher = {{Verein zur F{\"{o}}rderung des Open Access Publizierens in den Quantenwissenschaften}},
  volume = {3},
  pages = {163},
  month = jul,
  year = {2019}
}

@article{pokharel2018,
  title = {Demonstration of Fidelity Improvement Using Dynamical Decoupling with Superconducting Qubits},
  author = {Pokharel, Bibek and Anand, Namit and Fortman, Benjamin and Lidar, Daniel A.},
  journal = {Phys. Rev. Lett.},
  volume = {121},
  issue = {22},
  pages = {220502},
  numpages = {6},
  year = {2018},
  month = {Nov},
  publisher = {American Physical Society},
  doi = {10.1103/PhysRevLett.121.220502},
  url = {https://link.aps.org/doi/10.1103/PhysRevLett.121.220502}
}

@article{sutherland2024,
  title = {Passive dynamical decoupling of trapped-ion qubits and qudits},
  author = {Sutherland, R. T. and Erickson, S. D.},
  journal = {Phys. Rev. A},
  volume = {109},
  issue = {2},
  pages = {022620},
  numpages = {7},
  year = {2024},
  month = {Feb},
  publisher = {American Physical Society},
  doi = {10.1103/PhysRevA.109.022620},
  url = {https://link.aps.org/doi/10.1103/PhysRevA.109.022620}
}

@article{alistair2021,
author = {Alistair W. R. Smith  and Kiran E. Khosla  and Chris N. Self  and M. S. Kim },
title = {Qubit readout error mitigation with bit-flip averaging},
journal = {Science Advances},
volume = {7},
number = {47},
pages = {eabi8009},
year = {2021},
doi = {10.1126/sciadv.abi8009},
URL = {https://www.science.org/doi/abs/10.1126/sciadv.abi8009}
}

@article{nation2021,
  title = {Scalable Mitigation of Measurement Errors on Quantum Computers},
  author = {Nation, Paul D. and Kang, Hwajung and Sundaresan, Neereja and Gambetta, Jay M.},
  journal = {PRX Quantum},
  volume = {2},
  issue = {4},
  pages = {040326},
  numpages = {9},
  year = {2021},
  month = {Nov},
  publisher = {American Physical Society},
  doi = {10.1103/PRXQuantum.2.040326},
  url = {https://link.aps.org/doi/10.1103/PRXQuantum.2.040326}
}

@article{bravyi2021,
  title = {Mitigating measurement errors in multiqubit experiments},
  author = {Bravyi, Sergey and Sheldon, Sarah and Kandala, Abhinav and Mckay, David C. and Gambetta, Jay M.},
  journal = {Phys. Rev. A},
  volume = {103},
  issue = {4},
  pages = {042605},
  numpages = {12},
  year = {2021},
  month = {Apr},
  publisher = {American Physical Society},
  doi = {10.1103/PhysRevA.103.042605},
  url = {https://link.aps.org/doi/10.1103/PhysRevA.103.042605}
}

@article{funcke2022,
  title = {Measurement error mitigation in quantum computers through classical bit-flip correction},
  author = {Funcke, Lena and Hartung, Tobias and Jansen, Karl and K\"uhn, Stefan and Stornati, Paolo and Wang, Xiaoyang},
  journal = {Phys. Rev. A},
  volume = {105},
  issue = {6},
  pages = {062404},
  numpages = {24},
  year = {2022},
  month = {Jun},
  publisher = {American Physical Society},
  doi = {10.1103/PhysRevA.105.062404},
  url = {https://link.aps.org/doi/10.1103/PhysRevA.105.062404}
}

@article{pokharel2024,
  title = {Scalable measurement error mitigation via iterative bayesian unfolding},
  author = {Pokharel, Bibek and Srinivasan, Siddarth and Quiroz, Gregory and Boots, Byron},
  journal = {Phys. Rev. Res.},
  volume = {6},
  issue = {1},
  pages = {013187},
  numpages = {13},
  year = {2024},
  month = {Feb},
  publisher = {American Physical Society},
  doi = {10.1103/PhysRevResearch.6.013187},
  url = {https://link.aps.org/doi/10.1103/PhysRevResearch.6.013187}
}

@misc{aharonov2025importance,
      title={{On the Importance of Error Mitigation for Quantum Computation}}, 
      author={Dorit Aharonov and Ori Alberton and Itai Arad and Yosi Atia and Eyal Bairey and Zvika Brakerski and Itsik Cohen and Omri Golan and Ilya Gurwich and Oded Kenneth and Eyal Leviatan and Netanel H. Lindner and Ron Aharon Melcer and Adiel Meyer and Gili Schul and Maor Shutman},
      year={2025},
      eprint={2503.17243},
      archivePrefix={arXiv},
      primaryClass={quant-ph},
      url={https://arxiv.org/abs/2503.17243}, 
}

@misc{aharonov2025reliable,
      title={Reliable high-accuracy error mitigation for utility-scale quantum circuits}, 
      author={Dorit Aharonov and Ori Alberton and Itai Arad and Yosi Atia and Eyal Bairey and Matan Ben Dov and Asaf Berkovitch and Zvika Brakerski and Itsik Cohen and Eran Fuchs and Omri Golan and Or Golan and Barak D. Gur and Ilya Gurwich and Avieli Haber and Rotem Haber and Dorri Halbertal and Yaron Itkin and Barak A. Katzir and Oded Kenneth and Shlomi Kotler and Roei Levi and Eyal Leviatan and Yotam Y. Lifshitz and Adi Ludmer and Shlomi Matityahu and Ron Aharon Melcer and Adiel Meyer and Omrie Ovdat and Aviad Panahi and Gil Ron and Ittai Rubinstein and Gili Schul and Tali Shnaider and Maor Shutman and Asif Sinay and Tasneem Watad and Assaf Zubida and Netanel H. Lindner},
      year={2025},
      eprint={2508.10997},
      archivePrefix={arXiv},
      primaryClass={quant-ph},
      url={https://arxiv.org/abs/2508.10997}, 
}

@article{lee2023,
  title = {Error Suppression for Arbitrary-Size Black Box Quantum Operations},
  author = {Lee, Gideon and Hann, Connor T. and Puri, Shruti and Girvin, S. M. and Jiang, Liang},
  journal = {Phys. Rev. Lett.},
  volume = {131},
  issue = {19},
  pages = {190601},
  numpages = {7},
  year = {2023},
  month = {Nov},
  publisher = {American Physical Society},
  doi = {10.1103/PhysRevLett.131.190601},
  url = {https://link.aps.org/doi/10.1103/PhysRevLett.131.190601}
}

@article{Berry2015,
   title={{Simulating Hamiltonian Dynamics with a Truncated Taylor Series}},
   volume={114},
   ISSN={1079-7114},
   url={http://dx.doi.org/10.1103/PhysRevLett.114.090502},
   DOI={10.1103/physrevlett.114.090502},
   number={9},
   journal={Physical Review Letters},
   publisher={American Physical Society (APS)},
   author={Berry, Dominic W. and Childs, Andrew M. and Cleve, Richard and Kothari, Robin and Somma, Rolando D.},
   year={2015},
   month=mar }

@article{Kieferov2019,
   title={{Simulating the dynamics of time-dependent Hamiltonians with a truncated Dyson series}},
   volume={99},
   ISSN={2469-9934},
   url={http://dx.doi.org/10.1103/PhysRevA.99.042314},
   DOI={10.1103/physreva.99.042314},
   number={4},
   journal={Physical Review A},
   publisher={American Physical Society (APS)},
   author={Kieferov\'a, M\'aria and Scherer, Artur and Berry, Dominic W.},
   year={2019},
   month=apr }

@article{Granet2024,
   title={Hamiltonian dynamics on digital quantum computers without discretization error},
   volume={10},
   ISSN={2056-6387},
   url={http://dx.doi.org/10.1038/s41534-024-00877-y},
   DOI={10.1038/s41534-024-00877-y},
   number={1},
   journal={npj Quantum Information},
   publisher={Springer Science and Business Media LLC},
   author={Granet, Etienne and Dreyer, Henrik},
   year={2024},
   month=sep }

@misc{Abanin2025,
      title={Constructive interference at the edge of quantum ergodic dynamics}, 
      author={Dmitry A. Abanin and others},
      year={2025},
      eprint={2506.10191},
      archivePrefix={arXiv},
      primaryClass={quant-ph},
      url={https://arxiv.org/abs/2506.10191}, 
}

@misc{Zhang2025,
      title={Quantum computation of molecular geometry via many-body nuclear spin echoes}, 
      author={C. Zhang and others},
      year={2025},
      eprint={2510.19550},
      archivePrefix={arXiv},
      primaryClass={quant-ph},
      url={https://arxiv.org/abs/2510.19550}, 
}

@misc{huang2025,
      title={{Fullqubit alchemist: Quantum algorithm for alchemical free energy calculations}}, 
      author={Po-Wei Huang and others},
      year={2025},
      eprint={2508.16719},
      archivePrefix={arXiv},
      primaryClass={quant-ph},
      url={https://arxiv.org/abs/2508.16719}, 
}

\appendix

\onecolumngrid
\pagebreak
\newpage

\section{Appendix: Spin system matrices}
\label{appendix:spinmatrix}

We present in this appendix the three spin system matrices we have employed for our Hamiltonian simulations. For the sake of clarity, we will decompose the matrices in sub-matrices and highlight in bold the chemical shift (in ppm), leaving the spin-spin $J$ couplings (in Hz) in standard text. These matrices are taken directly from SPINACH examples.

\subsection{The DFH molecule}
The $16\times 16$ spin system matrix reads
\begin{equation}
\mathcal{M}^{\text{DFH}} =
\begin{pmatrix}
A_9 & B_{79}\\
0 & C_7
\end{pmatrix},
\end{equation}
with
%\begin{widetext}
\begin{equation*}
A_ 9 = \begin{pmatrix}
        \textbf{1.7970} & -15.1172 & 14.0478 & 10.0564 & 0 & 0 & 0 & 0 & 0 \\
        0 & \textbf{1.7970} & 38.0147 & 2.1802 & 0 & 0 & 0 & 0 & 0 \\
        0 & 0 & \textbf{-184.1865} & 49.6307 & 27.7945 & 18.4317 & 0 & 0 & 0 \\
        0 & 0 & 0 & \textbf{4.6834} & 4.5308 & 7.5739 & 0 & 0 & 0\\
        0 & 0 & 0 & 0 & \textbf{1.6370} & 0 & 7.4500 & 7.4500 & 7.4500 \\
        0 & 0 & 0 & 0 & 0 & \textbf{1.6942} & 7.4500 & 7.4500 & 7.4500 \\
        0 & 0 & 0 & 0 & 0 & 0 & \textbf{1.0092} & 0 & 0 \\
        0 & 0 & 0 & 0 & 0 & 0 & 0 & \textbf{1.0092} & 0 \\
        0 & 0 & 0 & 0 & 0 & 0 & 0 & 0 & \textbf{1.0092} \\
\end{pmatrix},
\end{equation*}
\begin{equation*}
B_ {79} = \begin{pmatrix}
        2.1802 & 38.0147 & 0 & 0 & 0 & 0 & 0 \\
        10.0564 & 14.0478 & 0 & 0 & 0 & 0 & 0 \\
        0 & 1.2295 & 0 & 0 & 0 & 0 & 0 \\
        0 & 0 & 0 & 0 & 0 & 0 & 0 \\
        0 & 0 & 0 & 0 & 0 & 0 & 0 \\
        0 & 0 & 0 & 0 & 0 & 0 & 0 \\
        0 & 0 & 0 & 0 & 0 & 0 & 0 \\
        0 & 0 & 0 & 0 & 0 & 0 & 0 \\
        0 & 0 & 0 & 0 & 0 & 0 & 0 \\
\end{pmatrix},
\end{equation*}
and
\begin{equation*}
C_{7} = \begin{pmatrix}
\textbf{4.6834} & 49.6307 & 4.5308 & 7.5739 & 0 & 0 & 0 \\
0 & \textbf{-184.1865} & 27.7945 & 18.4317 & 0 & 0 & 0 \\
0 & 0 & \textbf{1.6370} & -14.4404 & 7.4500 & 7.4500 & 7.4500 \\
0 & 0 & 0 & \textbf{1.6942} & 7.4500 & 7.4500 & 7.4500 \\
0 & 0 & 0 & 0 & \textbf{1.0092} & 0 & 0 \\
0 & 0 & 0 & 0 & 0 & \textbf{1.0092} & 0 \\
0 & 0 & 0 & 0 & 0 & 0 & \textbf{1.0092} \\
\end{pmatrix}.
\end{equation*}
%\end{widetext}

\subsection{The symmetric system with two coupled tert-butyl groups}
For the simulations of symm\_H and symm\_P the $22\times 22$ spin system matrix reads

\begin{equation}
    \mathcal{M}^{\text{symm}} = 
    \begin{pmatrix}
    A_{2} & B_{2,20}\\
    0 & C_{20}
    \end{pmatrix},
\end{equation}
with
\begin{equation*}
    A_2 = 
    \begin{pmatrix}
        \textbf{-43.844} & 301.990\\
        0 & \textbf{-43.844} 
    \end{pmatrix},
\end{equation*}
$C_{20}$ as a $20\times 20$ diagonal matrix defined as
\begin{equation*}
    C_{20} =
    \begin{pmatrix}
        \textbf{4.090} & & & & &\\
        & \textbf{4.090} & & & &\\
         & & \textbf{1.354} & & & \\
        & & & \ddots & & \\
        & & & & & \textbf{1.354}        
    \end{pmatrix},
\end{equation*}
and we decompose the matrix $B_{2,20}$ as
\begin{equation*}
    B_{2,20} =
    \begin{pmatrix}
        B_2 & B_{2,18}
    \end{pmatrix},
\end{equation*}
where
\begin{equation*}
    B_{2} =
    \begin{pmatrix}
        -321.62 & -19.15\\
        -19.15 & -321.62
    \end{pmatrix}
\end{equation*}
and
\begin{equation*}
    B_{2,18} = \Big(
    \overbrace{\begin{matrix}
        15.63 & \cdots & 15.63\\
        0 & \cdots & 0
    \end{matrix}}^{9\times}
    \overbrace{\begin{matrix}
        0 & \cdots & 0\\
        15.63 & \cdots & 15.63
    \end{matrix}}^{9\times}
    \Big).
\end{equation*}

\subsection{The phosophorous cluster}
For the phosphorous cluster the $34\times 34$ spin system matrix reads
\begin{equation}
    \mathcal{M}^{\text{phosphorous}} = \begin{pmatrix}
        A_7 & B_{7,27}\\
        0 & C_{27},
    \end{pmatrix}
\end{equation}
with\vspace{1cm}
%\begin{widetext}
\begin{equation*}
    A_7 = \begin{pmatrix}
        \textbf{-99.60} & -323.22 & -323.22 & -323.22 & 46.18 & 46.18 & 46.18\\
        0 & \textbf{-0.33} & -16.63 & -16.63 & -354.82 & 25.79 & -9.04\\
        0 & 0 & \textbf{-0.33} & -16.63 & -9.04 & -354.82 & 25.79\\
        0 & 0 & 0 & \textbf{-0.33} & 25.79 & -9.04 & -354.82\\
        0 & 0 & 0 & 0 & \textbf{-156.70} & -214.10 & -214.10\\
        0 & 0 & 0 & 0 & 0 & \textbf{-156.70} & -214.10\\
        0 & 0 & 0 & 0 & 0 & 0 & \textbf{-156.70}
    \end{pmatrix},
\end{equation*}
%\end{widetext}
$C_{27}$ a $27\times 27$ diagonal matrix defined as
\begin{equation*}
    C_{27} = \begin{pmatrix}
        \textbf{0.21} & & \\
        & \ddots & \\
         & & \textbf{0.21}
    \end{pmatrix},
\end{equation*}
and finally the matrix $B_{7,27}$ defined as
%\begin{widetext}
    \begin{equation*}
        B_{7,27} = \overbrace{\begin{pmatrix}
            0 & \cdots & 0 & 0 & \cdots & 0 & 0 & \cdots & 0\\
            4.0 & \cdots & 4.0 & 0 & \cdots & 0 & 0 & \cdots & 0\\
            0 & \cdots & 0 & 4.0 & \cdots & 4.0 & 0 & \cdots & 0\\
            0 & \cdots & 0 & 0 & \cdots & 0 & 4.0 & \cdots & 4.0\\
            0 & \cdots & 0 & 0 & \cdots & 0 & 0 & \cdots & 0\\
            0 & \cdots & 0 & 0 & \cdots & 0 & 0 & \cdots & 0\\
            0 & \cdots & 0 & 0 & \cdots & 0 & 0 & \cdots & 0
        \end{pmatrix}}^{\text{3 blocks of 9 numbers}}
    \end{equation*}
%\end{widetext}

\section{Appendix: Protocol for defining the number of shots in the \texorpdfstring{\ionq\ }{Forte Enterprise} experiment}
\label{appendix:ionqprotocol}

\begin{figure}[!ht]
    \centering
    \subfloat[Random 20 time points]{
    \includegraphics[scale=0.4]{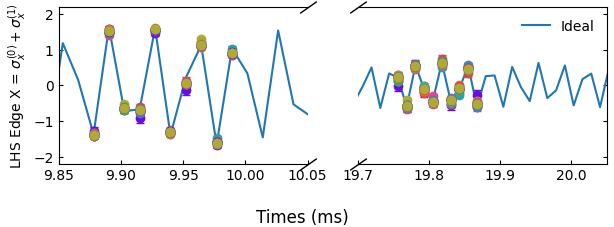}
    \label{fig:ionq-protocol-left}}
    \subfloat[MSE over the 20 time points as a function of $n_{shots}$]{
    \includegraphics[scale=0.4]{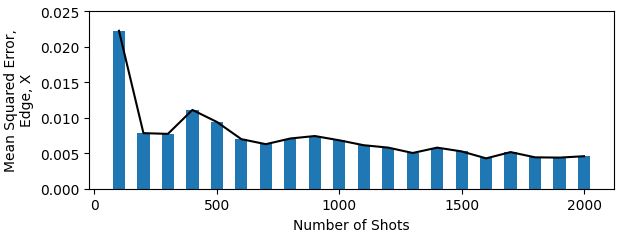}
    \label{fig:ionq-protocol-right}}
    \caption{Protocol to define the optimal number of shots $n_{shots}$ on \ionq\ for the simulation of the $^{31}\text{P}$ symm\_P spectrum. Left: A random selection of 20 times points for the $X$ magnetization as a function of $n_{shots}$. The blue points correspond to $n_{shots}=100$, the other colors correspond to all the other choices of $n_{shots}$ up to $n_{shots}=2,000$ in steps of 100. Right: The mean square error over the sum of the 20 random time points for the magnetization $X$ as a function of $n_{shots}$.}
    \label{fig:ionq-protocol}
\end{figure}

To define the optimal number of shots we randomly selected 20 points spanning the time interval and performed several runs for the $X$ magnetization only, as the results would be similar for the $Y$ magnetization. We evaluated the circuits for $n_{shots} \in [100:2,000]$ in steps of 100, and compared the output to the ideal simulated noiseless case using the local qiskit {\tt aer} simulator, as shown in Figure~\ref{fig:ionq-protocol} (left). Each color is associated with a different value of $n_{shots}$, the blue points corresponding to $n_{shots}=100$ with the largest error. We calculated the mean square error (MSE) of the 20 time points comparing the computation on \ionq\ with the noiseless simulation. The results are shown in Figure~\ref{fig:ionq-protocol} (right) and show an approximate plateau of the MSE at $n_{shots}=700$: from that value, increasing the number of shots only marginally increases the accuracy of the computation. We have thus selected this value of 700 shots for the full NMR spectrum simulation.

\section{Appendix: FID for the $^{31}$P 1D NMR spectrum of the symmetric P-H molecule calculated with the IonQ \texorpdfstring{\ionq\ }{Forte Enterprise} quantum computer}
\label{appendix:fidionq}

\begin{figure}[!ht]
    \centering
    \includegraphics[scale=0.8]{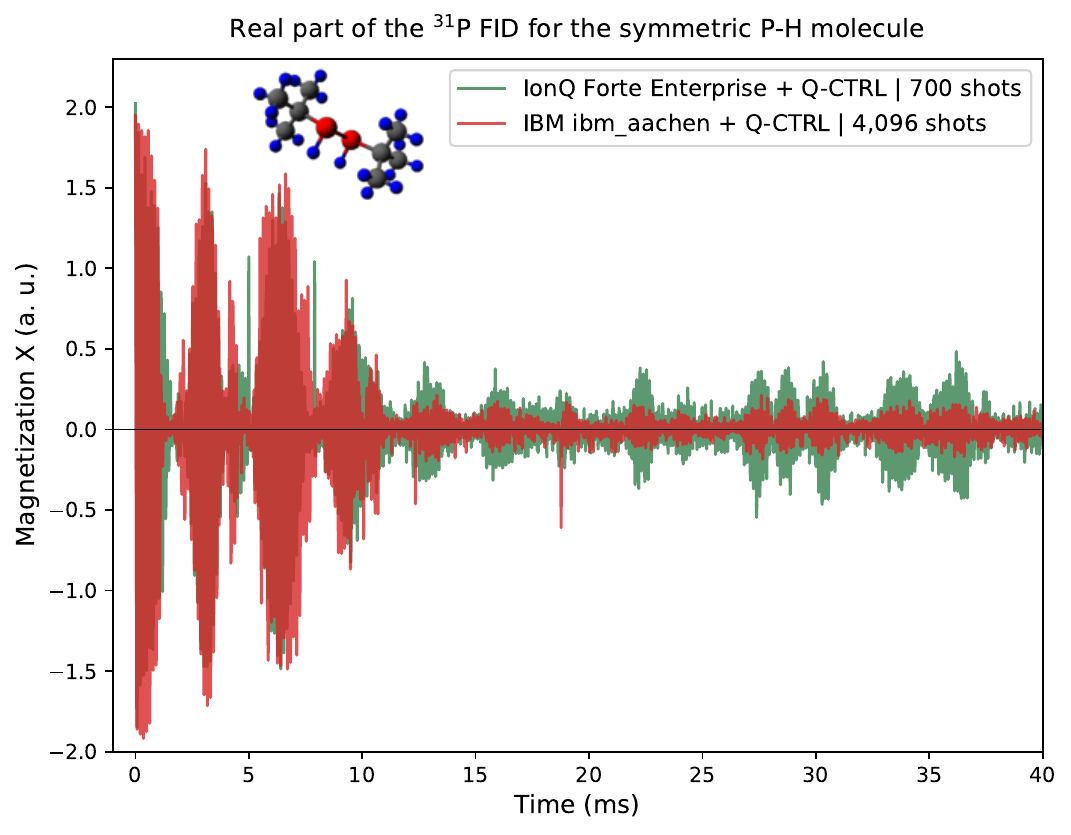}
    \caption{Real part of the FID ($X$ magnetization) for the symm P system as a function of time (in milliseconds). The red line stands for the computation using IBM \ibm\ hardware, with $4,096$ shots, while the green line stands for the computation using IonQ \ionq\ hardware, with $700$ shots. Both are using the Q-CTRL error suppression and mitigation pipeline.}
    \label{fig:fidionq}
\end{figure}

We present the results for the quantum computation of the FID for the symm\_P system ($^{31}$P spectrum of the symmetric P-H molecule), as obtained with IonQ \ionq\ and compared to the computation on \ibm. We display results for the real part of the FID only ($X$ magnetization), since the conclusions would be identical for the imaginary part ($Y$ magnetization). As seen in Figure~\ref{fig:fidionq} the two computations produce similar results, but the IonQ results are smoother and some outliers noticeable in the IBM computation, such as around $t=19$~ms, are absent. 

\section{Appendix: Comparison of our simulation to a noiseless simulation of quantum circuits}
\label{appendix:noisevsalgo}

\begin{figure}[!ht]
    \centering
    \includegraphics[width=\textwidth]{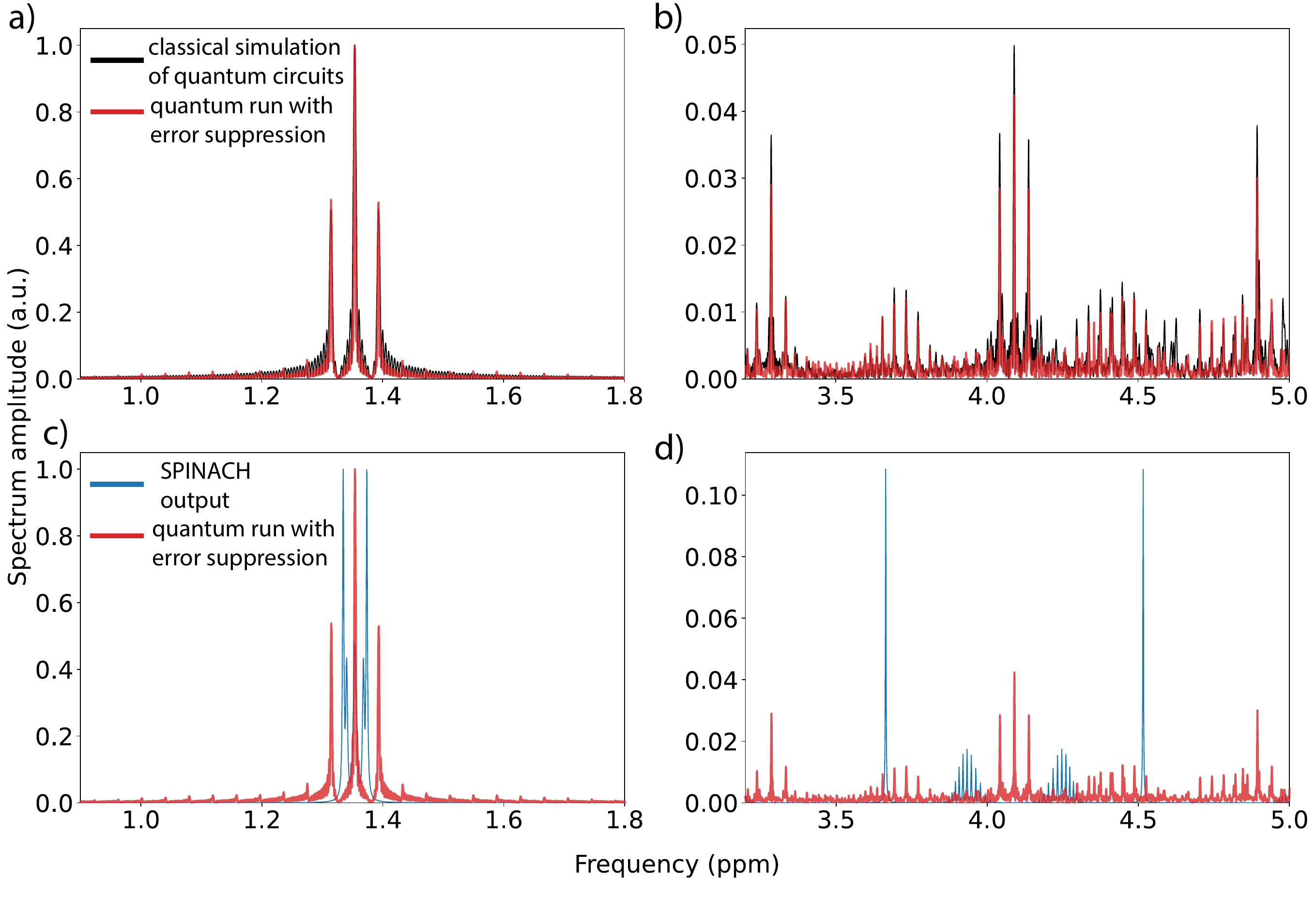}
    \caption{a) and b): Classical noiseless simulation of quantum circuits (black lines) compared to our computation on IBM \ibm\ user the error suppression and mitigation pipeline (red lines). The MSE is $8.0\times 10^{-5}$ and the cosine similarity is $0.9011$. c) and d): Classical computation using SPINACH (blue lines) compared to our computation (red lines). The MSE is $0.00109$ and the cosine similarity is $0.3596$.}
    \label{fig:noisevsalgo}
\end{figure}

We show significant regions of the spectrum of symm\_H simulated by our pipeline, compared to a spectrum obtained by an ideal (noiseless) classical simulation of the corresponding quantum circuits (see Figure \ref{fig:noisevsalgo} a), b)) using qiskit aer simulator, as well as to a spectrum obtained from SPINACH (see Figure \ref{fig:noisevsalgo} c), d)). The MSE is $8.06\times 10^{-5}$ and the cosine similarity is $0.9011$ for the classical simulation of quantum circuits compared to our computation on IBM \ibm\ using the error suppression and mitigation pipeline. When comparing the SPINACH computation to our computation, we obtain an MSE of $0.00109$ and a cosine similarity of $0.3596$. From the MSE and cosine similarity metrics, as well as by simply looking at the spectra, it can be clearly stated that the quantum hardware noise is negligible compared to the algorithmic error for this context and it supports our future direction where other methods providing a better approximation of the full Hamiltonian evolution combined with our error suppression and mitigation pipeline are expected to bring us even closer to quantum utility.

\end{document}